# "Interaction between the jet and the interstellar medium of M87 using *Chandra*"

S.Osone
Funabashi, Chiba, Japan, 273-0865.
Phone number :+81-47-422-6911, e-mail : osonesatoko@gmail.com

Abstract

This study shows the existence of an additional thermal component found in the synchrotron emission from the M87 jet. Using *Chandra*, thermal bremsstrahlung is detected in four out of six fields. X-rays for the nucleus and HST-1 are well described with a power law model. Neutral density, plasma density, and the density of accelerated electrons for the nucleus and each part of the jet are obtained. The mass of the thermal gas is estimated to be $3.9 \times 10^6$ $M_\odot$. The possibility that the thermal gas is injected into the jet from the black-hole is excluded because the nucleus and the field HST-1 do not contain thermal components. It is possible that the interstellar medium is compressed by the jet and is heated by the shock. The energy density of the gas is comparable to that of the magnetic field with 1 mG. For the field D, a high density of accelerated electrons of $1.5 \times 10^{11}$ ($2 d / t_{val}$)$^3$ ( $6 / \delta$ )$^3$ ( $3 / \Gamma$ )$^2$ m$^{-3}$ GeV$^{p-1}$, a high upper limit of neutral density of 8 cm$^{-3}$, and a high ion density of 195 cm$^{-3}$ are obtained. A possible contribution of non-thermal bremsstrahlung from the jet, especially the field D to the observed flux with *Fermi* is suggested. This scenario matches with non-flux variability in the GeV energy range.

1. Introduction

M87 is a close radio galaxy and is the center of the Virgo cluster. The distance is 16 Mpc (z=0.004) (Tonry 1991). The mass of the black hole is estimated to be $(3\sim6) \times 10^9$ $M_\odot$ (Macchetto et al. 1997, Gebhardt & Thomas 2009).

M87 is observed over a wide spectrum from radio frequencies to TeV gamma rays. The TeV gamma ray emissions from M87 were discovered by HEGRA(Aharonian et al. 2003) and confirmed with HESS, VERITAS and MAGIC (Aharonian et al. 2006, Acciari et al. 2008, Albert et al. 2008). M87 has an inclined jet of 20" length. The jet is resolved in radio with an angular resolution of 0".4, optical with an angular resolution of 0".7 and X-ray with an angular resolution of 0".7 of *Chandra*. The jet cannot be resolved in the high-energy range because *Fermi* has an angular resolution of 5'.2 and the Cherenkov image in TeV gamma rays has an angular resolution of 6'. Therefore, the origin of TeV gamma rays is studied with both, a timescale of flux variability and a correlation of flux with simultaneous observation. Flux variability of a timescale $t_{val}$ of 2 d in TeV gamma rays was observed (Aharonian et al. 2006). The size of the emitting area is given by $ct_{val}\delta = 3.1 \times 10^{16}(\delta/6)$ cm (0.01 pc). Here, $c$ is the speed of light and $\delta$ is the Doppler factor. Therefore, they conclude that the origin of TeV gamma rays is the nucleus. A correlation between a nucleus with radio of 43 GHz and TeV gamma rays with VERITAS also shows the same result (Acciari et al. 2009). A correlation between X-rays of the nucleus and TeV gamma rays was detected in 2008 and 2010, while that between X-rays of a field HST-1 and TeV gamma rays was detected in 2005 (Abramowaski et al. 2012). The field HST-1 is also an origin of TeV gamma rays. Radio, optical, and X-ray observations of the nucleus are used for the study of multi-wavelength energy spectra.

Non-simultaneous multi-wavelength energy spectra are described with various models (Reiger & Aharonian 2012). The models are classified into three categories: leptonic models, hadronic models and others. A leptonic model can be a homogeneous one-zone synchrotron self Compton model (Finke, Dermer and Bottcher 2008), decelerating jet model (Georganopoulus, Perlman and Kazanas, 2005), an electron of jet model (Stawarz et al. 2005, 2006), a multi-blob model (Lenain et al 2008), a spine-sheath layer model (Tavecchio & Ghisellini, 2008), a jet in a jet model (Giannios, Uzdensky and Begelman 2010) and an external inverse Compton model (Cui et al. 2012). A hadronic model can be a proton synchrotron model (Reimer, Protheroe and Donea, 2004), an interaction model between protons and an interstellar medium (Pfrommer & Enβlin 2003), and an interaction between protons and a cloud injected to the jet(Barkov, Ramon and Aharonian 2012). Others include a lepto-hadronic model (Reynoso, Medina and Romero 2011), dark matter annihilation model(Baltz et al, 2000, Saxena et al 2011) and magnetospheric model (Levinson & Rieger 2011).

NANTEN found a monocular cloud along an X-ray jet for SS433 (Yamamoto et al., 2008). The closest cloud N4 is 20 pc from SS433. The CO density for N1 is 3 cm$^{-3}$. They insist that the jet compresses the interstellar medium and a monocular cloud is formed. The interaction between the interstellar medium and an electron is expected. For the north part of SS433, which matches with the monocular cloud N4, the energy spectra are well described with either a power law with a photon index of 1.88 or thermal bremsstrahlung of 6 keV (Moldowan et al 2005). For the south part of SS433, which does not match with any monocular cloud, energy spectra are well described with a combination of a power law with a photon index of 2.41 and a Mekal model with 0.20 keV (Brinkmann et al. 2007) .

Recently, the detection of diffused thermal component of 1 keV for a radio lobe of the radio galaxies Cen A (Stawarz et al. 2013, O'sullivan et al. 2013) and Fornax A(Seta, Tashiro and Inoue, 2013) was reported. They suggest that the gas was not ejected from the central engine but was transported into the jet from surroundings.

There is the possibility of a thermal component for not only the radio lobe of the jet, but also the jet itself. The jet of M87 has been analyzed with *Chandra* (Wilson & Yang 2002, Perlman & Wilson 2005). However, all parts of the jet are fitted with a power law. The jet of M87 has also been analyzed with *XMM* which could resolve only the outer part of the jet (Bohringer et al. 2001). The energy spectra are well described with a power law and they conclude that no thermal component is present in the outer part of the M87 jet.

In this thesis, the first such study of the thermal component in the X-ray spectra for each part of M87 jet using *Chandra* is presented. The flux of both non-thermal bremsstrahlung and gamma rays from pion decay in GeV energy range for each part of the jet is presented.

2. Observation

Data is used for this study are archival data and are the same as those in (Wilson & Yang 2002, Perlman & Wilson 2005). The observation date is July 29, 2000 (ObsID 352) and July 30,2000 (obsID 1808). The frame time is 3.2 s for ObsID352 and 0.4 s for ObsID 1808. For ObsID 352, CCD I2, I3, S1, S2, S3, S4 were used. This was for the faint part of the jet. Good exposure time is 37.6 ks. For ObsID 1808, only S3 was used. This was for the bright part of the jet. Good exposure time is 12.8 ks. For the fields N (Nucleus), HST-1, D and A, data of ObsID 1808 were used, because it was reported that their observation saturated in a 3.2 s frame.

3. Result

The CIAO package 4.5 and the CALDB 4.5.6 downloaded from the *Chandra* Homepage were used for this study.

3.1 Images

Fig. 1 is the image of a jet extracted from Wilson & Yang (2002). The same position and radius of circle for each part of the jet in Wilson & Yang (2002) are used. The field HST-1 was not analyze in Wilson & Yang (2002). For the field HST-1, same position and radius as in Perlman & Wilson (2005) are used. Table 1 shows the position data of the extracted. The method of extraction of background is the same as that used in the study of M87 jet with *Chandra* (Wilson & Yang 2002, Perlman & Wilson 2005) and with *XMM* (Bohringer et al.2001). Two neighboring regions of same size, are taken as background region for each part of the jet as shown in Fig. 2. Hot gas of the Virgo cluster centered at M87 has been reported with *XMM* (Belsole et al. 2001). It is considered that no effect of thermal hot gas of the cluster on the data which subtract background. This correctness is checked in section 3.2.3. Peculiar X-ray emission outside the jet of M87 was pointed out between field E and field F (Dainotti et al 2012). Background for the field E, the field F and the field I cannot be estimated correctly. Therefore, these three fields are not analyzed in present study.

3.2 Energy spectra

In Wilson & Yang (2002) and Perlman & Wilson (2005), the energy spectra of the M87 jet are fitted with a power law model as synchrotron emission of an accelerated electron. The photo absorption of a cold interstellar medium is included in this model. However, it is interest to find a component of thermal bremsstrahlung, as an interaction between an electron and the interstellar medium compressed by the jet. Hence, two models are used, namely a power law, a combination of a power law and thermal bremsstrahlung.

3.2.1 Model fitting

CCD has a sensitivity from 0.2 keV to 10 keV. There is a quantum efficiency degradation by the contamination of an optical filter. The energy range above 0.3 keV, and especially above 0.5 keV is well calibrated. Therefore, an energy range of 0.5 keV to 5 keV is employed, where the 5 keV limit is a statistical limit. Data points are binned so that the minimum counts per bin are above 20. The XSPEC tool is used for model fitting, and following the method in Wilson & Yang (2002), the photo absorption is first fixed to $0.018 \times 10^{22}$ cm$^{-2}$ (Kalberia et al., 2005). After the first fit, the photo absorption is a free parameter and fit.

3.2.2 Comparison with previously published results

In Wilson & Yang (2002), *Chandra* data for M87 jet were analyzed. However, they did not consider the effect of a quantum efficiency degradation of the CCD, which was discoverd at later. In Perlman & Wilson (2005), *Chandra* data for the M87 jet were reanalyzed using an ACISABS program. The consistency of the fitting result that reported in Perlman & Wilson (2005) is checked. In Perlman & Wilson (2005), an energy range of (0.3, 5) keV was used. Therefore, data are prepared in the energy range of (0.3, 5 ) keV and fitted with a power law, with photo absorption included. Three fitting parameters, $N_H$, the photon index, and 1 keV flux for the field A are compared. Here, the photon index α is described with $dN/dE \propto E^{-\alpha}$. In Perlman & Wilson (2005), they obtain α of (2.61± 0.07), $N_H$ of (0.08±0.05) x$10^{22}$ cm$^{-2}$, 1 keV flux of 2.77x$10^{-4}$ ph/cm$^2$/s/keV, while in the present study, values of α of (2.43 ± 0.06 ), $N_H$ of (0.00+0.04) x $10^{22}$ cm$^{-2}$, 1 keV flux of (3.34 +0.09 −0.10) x$10^{-4}$ ph/cm$^2$/s/keV are obtained. The error is 90% confidence level statistical error. $N_H$ of Perlman & Wilson (2005) match closely with the results of present study, within an error range. Recent calibration data, which include a detector-position dependent calibration, are used in present analysis. A difference of photon index and 1 keV flux may be attributed to that.

3.2.3  Fitting result

From Fig.3 to Fig.8 show the fitted energy spectra for all fields, respectively. Table 2 shows the fitting results of the two models. An error could not be obtained for some of fitting values. The reduce χ$^2$ of the two models are compared, and detection of thermal bremsstrahlung is defined to exist when the combination model of a power law and thermal bremsstrahlung is better than the power law model. Thermal bremsstrahlung for four fields, D, A, B, C+G, among six are found. The closest field among these four is 220 pc from nucleus. The temperature of thermal bremsstrahlung ranges from 0.1 to 0.2 keV. The radius of extracted field for the field N and HST-1 is almost identical to PSF. The angular resolution is dependent on energy. The encircled energy is 0.7954 for 0.11 keV, 0.7817 for 0.52 keV and 0.6426 for 5.41 keV for 1" diameter. Small radius may favor low energy photons. However, the detection of thermal component is in the outer part of the jet with larger radius. The temperature is almost same with thermal component from the jet for SS433 of 0.20 keV (Brinkmann et al 2007), for For A of 1 keV (Seta, Tashiro and Inoue 2013) and for Cen A of 0.5 keV (Stawarz et al, 2013).

In the energy spectra for the field N, there is a line feature, which is considered as contamination of hot gas of the Virgo cluster. The background subtraction for the field N is failed because proper background region for the center region cannot be taken. However, good reduced χ$^2$ of 0.842 is obtained when the energy spectra is fitted with power law. Therefore, line contamination can be ignored statistically. The count of background region and source region are compared in order to check if the background subtraction is correct. Total count is 2470 cts for background 170 cts (exposure time 12.8 ks) for the field HST-1. The significance of source count is 180 sigma. Total count is 2300 cts for background 1200 cts (exposure time 37.6 ks) for the field C+G, which is most faint part of the jet. The significance of source count is 32 sigma. And, the temperature of hot gas of cluster is higher than 1 keV over 1 kpc of the jet (Bohringer et al. 2001). Therefore, possible contamination of the hot gas of the Virgo cluster can be neglected.

The flux (0.5-5 keV) of the field N is $8.39 \times 10^{-13}$ erg/s/cm². All $N_H$ are same with neutral column density from 21 cm measurement of $0.018 \times 10^{22}$ cm$^{-2}$ (Kalberla et al. 2005) within 90% confidence error.

4. Discussion

4.1 Physical values

Three physical values were calculated: upper limit of neutral density $N$ from $N_H$, plasma density $n_i\, n_e$ from a normalization of thermal bremsstrahlung, and the density of accelerated electrons $K'$ from 1 keV flux of a power law. Table 3 shows three physical values.

4.1.1 Neutral density

X-ray is absorbed by the neutral material in line of sight from us to M87. The upper limit of neutral density can be calculated with a diameter of extracted field. The upper limit of neutral density $N$ (cm$^{-3}$) is given as $N = N_H / 2R$. Here, $R$ is the radius of the extracted field.

4.1.2 Density of an accelerated electron

Synchrotron emissivity J is given by $2.344 \times 10^{-25} \times a(p)\, B^{(p+1)/2}\, K'\, (3.217 \times 10^{17}/\nu)^{(p-1)/2}$ W m$^{-3}$ Hz$^{-1}$ (Longair 1992). Here, the density of an accelerated electron is given by $dN_e/dE = K'\, E^{-p}$ m$^{-3}$ GeV$^{-1}$. $E$ is in units of GeV and $K'$ is in units of m$^{-3}$ GeV$^{p-1}$. $B$ is a magnetic field in units of T. Various values of the magnetic field are reported from different analyses. The one-zone synchrotron self Compton model with fitted multiwavelength energy spectra deduces the magnetic field to be 55 mG (Abdo et al, 2009). The spine-sheath layer model with fitted multiwavelength energy spectra deduces the magnetic field of 1 G (Aleksic et al. 2012). The X-ray of nucleus is used for multiwavelenght energy spectra. The observation of synchrotron self-absorption of the nucleus N at 43 GHz sets the limit on the magnetic field from 1 G to 10 G (Kino et al., 2014). The X-ray energy loss in the field HST-1 is consistent with an $E^2$ energy loss of synchrotron emission, and is dependent on the magnetic field. From this, the magnetic field 0.6 mG for $\delta=5$ is obtained (Harris et al. 2009). Variability in X-ray is different for different part of the jet, as the field N is variable on short timescale, the field HST-1 and the field D show similar variability and the field A is non variable over 10 years (Harris et al. 2009). Timescale of the energy loss of synchrotron emission is proportional to $1/B^2$ (Longair 1992). The magnetic field is considered to be different for different part of the jet. The magnetic field of 1 G is used for field N and 1 mG is used for other fields. The index $p$ is given by $2\alpha-1$ ($\alpha$ is the photon index). The numerical parameter $a(p)$ is a constant that depends on the index $p$. $a(3.5)$ is 0.217 and $a(4)$ is 0.186. $\nu$ is the frequency in units of Hz. The 1 keV flux of synchrotron emission is given by J $V_{jet}$ $\Gamma^2/4\pi D^2$. Here, $D$ is the distance of M87. Here, $\Gamma$ is the Lorentz factor. There is a beaming effect for accelerated electrons. The solid angle in observer system is $1/\Gamma^2$ times as large as that in the jet system. Here, $V_{jet}$ is the volume of the jet, which is given as $4\pi R_{jet}^3/3$. Also, $R_{jet}$ is $ct_{val}\delta$. $t_{val}$ is the timescale of flux variability in the observer system and $\delta$ is the Doppler factor given by $1/\Gamma(1-\beta\cos\theta)$. $\beta$ is given by $v/c$, $v$ is speed of the jet, $c$ is the speed of light and $\theta$ is inclination angle of the jet. Variability in the TeV energy range is reported to be 2 d (Aharonian et al. 2006). Variability in the X-ray range is reported to be a few days for the nucleus (Hariss et al 2011). The apparent velocity observed with the *Hubble Space Telescope* is $6c$ for a field HST-1, that means an inclination angle of $\theta=19°$ and a Lorentz factor of $\Gamma=3$, and an apparent velocity is different for each part of the jet (Biretta, Sparks and Macchetto 1999, Meyer et al. 2013). I derive $K$' with $t_{val}$ = 2 d, a Doppler factor $\delta$ =6, and a Lorentz factor $\Gamma=3$ for all parts of the jet. $V_{jet}$ is given as $V_{jet}$ = $1.2\times10^{50}$ ($t_{val}$/ 2 d )$^3$( $\delta$ / 6 )$^3$ cm$^3$. With longer timescale of flux variability, $V_{jet}$ is larger and $K$' is smaller.

4.1.3 Plasma density

The normalization of thermal bremsstrahlung is $3.02\times10^{-15}\times n_e n_i V/4\pi D^2$. Here, $n_e$ is the electron density in units of cm$^{-3}$ and $n_i$ is the ion density in units of cm$^{-3}$. $D$ is the distance of M87 in units of cm. $V$ is the volume of the extracted field in units of cm$^3$ (given as $4\pi R^3/3$), and $R$ is the radius of the extracted field in units of cm as shown in Table 1. $V$ for the field D is $2.4\times10^{61}$ cm$^3$. The electron density $n_e$ is assumed to be equal to the ion density $n_i$ and the ion density $n_i$ is derived. The calculated total mass of the thermal component is $M=m_H n_i V$, as shown in Table 4. Here, $m_H$ is the mass of a hydrogen. Mass of the thermal gas for a field D is small, $3.9\times10^6$ M$_\odot$, compared to the mass of the black hole, $(3\sim6)\times10^9$ M$_\odot$ (Macchetto et al 1997, Gebhardt & Thomas 2009). This result is different from For A and Cen A where the mass of the thermal gas was more than the mass of the black-hole by several orders (O'sullivan et al 2013, Seta,

Tashiro and Inoue, 2013). The possibility that the thermal gas is injected from the black-hole into the jet is excluded because the X-ray energy spectra for the nucleus and the field HST-1 do not contain thermal components. This argument is same with For A and Cen A because high mass of the thermal gas cannot be ejected from the black-hole (O'sullivan et al 2013, Seta, Tashiro and Inoue, 2013). They suggest the thermal gas is transported into the jet from surroundings. It is not discussed how to heat the material. NANTEN observation of CO along X-ray jet of SS433 suggests that the jet compress the interstellar medium (Yamamoto et al. 2008). Therefore, it is possible that the interstellar medium of surroundings is compressed by the jet and is heated by the shock. The description that an accelerated electron is surrounded by the thermal gas in the jet is suggested. The limb brighten cannot be observed because flux ratio of thermal bremsstrahlung to power law is almost 0.20.

The energy density of the magnetic field is given by $B^2/2\mu_0$. Here, $B$ is a magnetic field in units of T and $\mu_0$ is the permeability of free space. The energy density with 1 mG is $3.8\times10^{-8}$ erg/cm$^3$. The calculated energy density of the thermal component $2 n_e kT$ is shown in Table 4. The energy density of the thermal component for the field D is $6.9\times10^{-8}$ erg/cm$^3$. The energy density of thermal gas is comparable to that of magnetic field.

The rotation measure along the jet of 1.2 kpc is RM~200 rad m$^{-2}$ (Owen, Eilek and Keel, 1990). RM is given as $8.12\times10^3 n_e B L$ (Longair 1992). Here, $n_e$ is plasma density in units of m$^{-3}$, $B$ is a magnetic field in units of T and $L$ is a size of region in units of pc. Plasma density is given as $2.1\times10^{-3}$ cm$^{-3}$ with a magnetic field of 1 mG and a size of 120 pc for field D. This value differs from the result of X-ray energy spectra by several orders. The lower limit of 90% confidence level of plasma density is 32 cm$^{-3}$ for the field D. The reason may be smaller magnetic field than 1 mG.

4.2. Contribution to flux in GeV energy range

Some fields have a high density of the neutral medium and the ion density. This allows for flux contribution from events of accelerated particles (electrons and protons) interacting with the interstellar medium compressed by the jet. Such a flux would be observed in the GeV range with *Fermi*. At 1 GeV, the calculated flux of both the non-thermal bremsstrahlung of an accelerated electron, and the gamma rays arising from pion decay by the interaction between accelerated protons and the interstellar medium are shown in table 5.

4.2.1 Non-thermal bremsstrahlung

Flux of non-thermal bremsstrahlung for a cold material I is given as $bN \times K \epsilon^{-p}/(p-1)$ ph/m$^3$/s/J (Longair 1992). Here, b is $1.03 \times 10^{-21}$ m$^3$/s. $N$ is the neutral density in unit of m$^{-3}$. $\epsilon$ is the energy in units of MeV. The number density of the accelerated electron is given as $dN_e/dE = K(E)^{-p}$. $K$ is in units of m$^{-3}$ J$^{p-1}$. $E$ is in units of J. The relation between $K$ and $K'$ is given as $K' = (1.6 \times 10^{-10})^{p-1} K$. The total energy loss rate $dE/dt$ of the non-thermal bremsstrahlung for fully ionized materials differs from that for cold materials by a factor of 2 for $100 < \gamma < 10^5$ (Longair 1992). Therefore, the same formula is used for fully ionized materials. Flux is given by I $V_{jet} \Gamma^2 / 4\pi D^2$. Flux variability $t_{val}$[(Harris et al. 2009) and apparent velocity that define the $\Gamma$ and $\delta$ (Biretta, Sparks and Macchetto, 1999, Meyer et al. 2013) are different for each part of the jet. The calculated flux of non-thermal bremsstrahlung is independent of $\Gamma$ and $V_{jet}$ which includes $t_{val}$ and $\delta$ because $\Gamma$ and $V_{jet}$ are used in a calculation of $K'$ from 1 keV flux of a power law. The calculated flux of non-thermal bremsstrahlung is shown in Table 4. There is a large dependence on the index $p$ of the flux. The flux is $1.4 \times 10^{-9}$ ph/s/cm$^2$/GeV at 1 GeV with *Fermi* (Abdo et al 2009). The calculated flux for the jet, especially the filed D is 100% of observed flux with *Fermi*. The calculated energy spectra are given by $dN/dE = 4.2 \times 10^{-2} E^{-4}$ ph/s/cm$^2$/GeV. Here, $E$ is an energy in units of GeV. With smaller (larger) magnetic field, $K$ is larger (smaller) and flux of non-thermal bremsstrahlung increase (decrease).

90% confidence level lower limit of flux of non-thermal bremsstrahlung for field D is calculated. The lower limit of $K' = 2.2 \times 10^9$ m$^{-3}$ GeV$^{p-1}$ is given by the lower limits of both the photon index α =1.98 ($p$=3) and the 1 keV flux of the power law. The lower limit of the flux is given by lower limits of both $K'$, index $p$ and the material density of 33 cm$^{-3}$. The lower limit of flux of non-thermal bremsstrahlung at 1 GeV is $2.0 \times 10^{-13}$ ph/cm$^2$/s/GeV. A possible contribution of non-thermal bremsstrahlung from the jet, especially the filed D to the observed flux of with *Fermi* is suggested.

4.2.2 Gamma ray flux from pion decay

Gamma rays emerging from pion decay are calculated for each part of the jet. The flux extracted from figure in Dermer (1986) is scaled with proton flux and the medium density. The extracted gamma ray flux $I_0$ is $2 \times 10^{-26}$ ph/cm$^3$/s/GeV at 1GeV for proton energy spectra $dN_p/dE = 2.2 \times (E/\text{GeV})^{-2.75}$ /cm$^2$/s/GeV/str and with medium density of 1 cm$^{-3}$. The flux of accelerated electrons is given as product of the density of the accelerated electron $dN_e/dE = K'(E/\text{GeV})^{-p}$ and $c/4\pi$. The electron and proton density is assumed to be the same at 1 GeV. An example for the calculation for the field D is shown. With $K' = 1.5 \times 10^{11}$ m$^{-3}$ GeV$^{p-1}$, the proton flux is scaled with $1.6 \times 10^{14}$. The gamma ray flux with the medium density $N + n_i = 203$ cm$^{-3}$ is given by $I_0$ x (1.6x10$^{14}$) x (203) x $V_{jet}$ $\Gamma^2/4\pi D^2$. The calculated flux for each part of the jet is presented in Table 4. The gamma ray flux from pion decay is lower than the non-thermal bremsstrahlung by several orders of magnitude.

4.2.3 Timescale of flux variability

The photon index is 2.26±0.13 in GeV energy range (Abdo et al 2009) and 2.21 in TeV energy range (Aleksic et al 2012). Energy spectra from GeV to TeV are smoothly connected. A flux variability of 2 d is reported in the TeV energy range [5]. The flux variability in the GeV energy range is studied with 10 months data, but the probability of the detection of variability is 22% (Abdo et al 2009). Hence, flux variability has not been detected.

The time scale of non-thermal bremsstrahlung is given by $1 / (3.66 \times 10^{-22} N)$ s for a cold material (Longair 1992). Here, $N$ is the neutral medium density in units of m$^{-3}$. With $N = 8.3 \times 10^6$ m$^{-3}$ for the field D, the timescale is 1.0x10$^7$ yr in the jet system. The timescale is 1.7x10$^6$ (6 / $\delta$) yr in observer system. The time scale of non-thermal bremsstrahlung is given as $1 / (7.0 \times 10^{-23} n_i (\ln\gamma + 0.36))$ s for a fully ionized hydrogen plasma (Longair 1992). Here, $n_i$ is the ion density in units of m$^{-3}$. With $n_i = 195 \times 10^6$ m$^{-3}$ for the field D, the timescale is 2.9 x10$^5$ yr in the jet system. The timescale is 4.8x10$^4$ (6 / $\delta$) yr in observer system. This is a long timescale match with non-flux variability in the GeV energy range and may indicate a different origin for the GeV gamma rays and TeV gamma rays, which shows an inverse Compton of synchrotron emission.

5. Conclusion

Four fields that has thermal bremsstrahlung added to synchrotron emission is found with *Chandra*. The closest field among them is 220 pc from field N. On the other hand, field N favors a power law model. X-ray spectrum of field N is rightly used for multi-wavelength energy spectra. The mass of the thermal gas is estimated as 3.9x10$^6$ M$_\odot$. The possibility that thermal gas is injected to the jet from the black hole is rejected because field N and a field HST-1 do not contain thermal components. This result suggests that thermal gas is transported from surroundings of the jet as CenA and For A. It is possible that the interstellar medium is compressed by the jet and is heated by the shock. The energy density of the gas is comparable to that of the magnetic field. For the field D, high density of accelerated electron of 1.5x10$^{11}$ (2 d / $t_{val}$ )$^3$ ( 6/ $\delta$ )$^3$ ( 3 / $\Gamma$ )$^2$ m$^{-3}$ GeV$^{p-1}$, a high neutral density of 8 cm$^{-3}$ and a high ion density of 195 cm$^{-3}$ are obtained. However, the plasma density is different from the rotation measure by several orders. There is a possible contribution of non-thermal bremsstrahlung from the jet, especially the field D to the GeV energy spectra observed with *Fermi*. This energy spectra are described with $dN/dE = 4.2 \times 10^{-2} E^{-4}$ ph/s/cm²/GeV. $E$ is energy in units of GeV. The timescale of non-thermal bremsstrahlung is $4.8 \times 10^4$ yr. This scenario matches with the non-variability in the GeV energy range.

Acknowledgement

I thank Dr. J. Kataoka for my idea through discussion. I thank prof. Y. Fukui for NANTEN observation result. I thank prof. K. Makishima for some comments about total analysis. I thank Dr. K.Matsushita for suggestion about background.

.

Table1. The extracted field for each part of the jet. 1" is 78 pc.

| Name | RA | DEC | Distance from N | radius of Region |
|---|---|---|---|---|
| N | $12^h 30^m 49^s.40$ | $12° 23' 27".8$ | | 0".5 |
| HST-1 | $12^h 30^m 49^s.36$ | $12° 23' 28".3$ | 0".8 (62 pc) | 0".6 |
| D | $12^h 30^m 49^s.23$ | $12° 23' 28".8$ | 2".8 (220 pc) | 0".75 |
| A | $12^h 30^m 48^s.60$ | $12° 23' 32".2$ | 12".4 (970 pc) | 1".0 |
| B | $12^h 30^m 48^s.51$ | $12° 23' 32".9$ | 14".2 (1100 pc) | 1".25 |
| C+G | $12^h 30^m 48^s.25$ | $12° 23' 34".8$ | 18".7 (1500 pc) | 1".7 |

Table2 The Results fitted for each part of the jet. Here, "PL" is a power law. "Bremss" is thermal bremsstralung. Photo absorption is included. Photon index α is defined as d$N$/d$E \propto E^{-\alpha}$. The error is a 90% confidence level statistical error.

| Name | | N | HST-1 | D |
|---|---|---|---|---|
| PL | $N_H$ (x $10^{22}$ cm$^{-2}$) | $0.05^{+0.04}_{-0.03}$ | $0.02^{+0.03}_{-0.02}$ | $0.00^{+0.03}$ |
| | photon index α | $2.16^{+0.17}_{-0.15}$ | $2.16^{+0.18}_{-0.15}$ | $2.06^{+0.18}_{-0.11}$ |
| | 1 keV flux (ph/cm$^2$/s/keV) | $(2.72^{+0.34}_{-0.30})$x$10^{-4}$ | $(2.08^{+0.29}_{-0.18})$x$10^{-4}$ | $(1.05^{+0.13}_{-0.06})$x$10^{-4}$ |
| | $\chi^2$/d.o.f | 0.842 | 1.263 | 0.645 |
| PL+Bremss | $N_H$ (x $10^{22}$ cm$^{-2}$) | $0.05^{+0.04}_{-0.02}$ | $0.11^{+0.25}_{-0.11}$ | $0.30^{+0.25}_{-0.28}$ |
| | photon index α | $2.16^{+0.17}_{-0.15}$ | $2.24^{+0.30}_{-0.44}$ | $2.45^{+0.51}_{-0.47}$ |
| | 1 keV flux (ph/cm$^2$/s/keV) | $(2.72^{+0.34}_{-0.30})$x$10^{-4}$ | 2.34x$10^{-4}$ | $(1.67^{+0.94}_{-0.70})$x$10^{-4}$ |
| | $kT$(keV) | 3.78x$10^{-3}$ | 0.17 | $0.11^{+0.05}_{-0.09}$ |
| | normalization | 4.96x$10^{-6}$ | 3.49x$10^{-3}$ | $(9.17^{+193}_{-8.92})$x$10^{-2}$ |
| | flux ratio (Bremss/total) | 0.00 | 0.10 | 0.22 |
| | $\chi^2$/d.o.f | 0.866 | 1.292 | 0.484 |
| Name | | A | B | C+G |
| PL | $N_H$ (x $10^{22}$ cm$^{-2}$) | $0.00^{+0.65}$ | $0.00^{+0.04}$ | $0.00^{+0.07}$ |
| | photon index α | $2.44^{+0.08}_{-0.09}$ | $2.18^{+0.22}_{-0.10}$ | $1.94^{+0.37}_{-0.15}$ |
| | 1 keV flux (ph/cm$^2$/s/keV) | $(3.35^{+0.12}_{-011})$x$10^{-4}$ | $(5.54^{+0.90}_{-0.26})$x$10^{-5}$ | $(2.61^{+0.77}_{-0.22})$x$10^{-5}$ |
| | $\chi^2$/d.o.f | 1.213 | 0.868 | 0.873 |
| PL+Bremss | $N_H$ (x $10^{22}$ cm$^{-2}$) | $0.05^{+0.15}_{-0.05}$ | $0.17^{+0.23}_{-0.17}$ | $0.29^{+0.44}_{-0.29}$ |
| | photon index α | $2.22^{+0.24}_{-0.23}$ | $2.24^{+0.39}_{-0.35}$ | $2.27^{+0.40}_{-0.42}$ |
| | 1 keV flux (ph/cm$^2$/s/keV) | $(3.12^{+0.85}_{-0.60})$x$10^{-4}$ | $(6.25^{+921}_{-2.03})$x$10^{-5}$ | $(3.88^{+4.25}_{-2.38})$x$10^{-5}$ |
| | $kT$(keV) | 0.15±0.09 | $0.18^{+0.21}_{-0.18}$ | $0.13^{+0.58}_{-0.13}$ |
| | normalization | $(9.81^{+73}_{-8.33})$x$10^{-3}$ | $(2.31^{+0.00}_{-2.30})$x$10^{-3}$ | $(7.67^{+2090}_{-7.67})$x$10^{-3}$ |
| | flux ratio(Bremss/total) | 0.15 | 0.22 | 0.19 |
| | $\chi^2$/d.o.f | 0.983 | 0.854 | 0.859 |

Table 3. Comparison of physics values. Here, "PL" is a power law. "Bremss" is thermal bremsstralung. ( )a is the numerical value used for a calculation of $K$'.

| name | Best Model | Neutral density $N$ (cm$^{-3}$) | plasma density $n_e n_i$ (cm$^{-6}$) | $p=2\alpha-1()^a$ | $K'$(m$^{-3}$GeV$^{p-1}$) |
|---|---|---|---|---|---|
| N | PL | 2.1 | | 3.32(3.5) | 7.1x10$^4$ |
| HST-1 | PL | 0.6 | | 3.32(3.5) | 3.0x10$^{11}$ |
| D | PL+Bremss | 8.3 | 3.8x10$^4$ | 3.9(4.0) | 1.5x10$^{11}$ |
| A | PL+Bremss | 1.1 | 1.7x10$^3$ | 3.44(3.5) | 4.4x10$^{11}$ |
| B | PL+Bremss | 2.8 | 2.1x10$^2$ | 3.48(3.5) | 9.0x10$^{10}$ |
| C+G | PL+Bremss | 3.5 | 2.7x10$^2$ | 3.54(3.5) | 5.6x10$^{10}$ |

Table4. List of physical values of the thermal component. The ion density and the thermal electron density are assumed to be equivalent.

| Name | ion density $n_i$(cm$^{-3}$) | $kT$(keV) | Volume $V$(cm$^3$) | Mass(M$_\odot$) | thermal gas pressure $p_T$(erg/cm$^3$) |
|---|---|---|---|---|---|
| D | 195 | 0.11 | 2.4x10$^{61}$ | 3.9x10$^6$ | 6.9x10$^{-8}$ |
| A | 41 | 0.15 | 5.7x10$^{61}$ | 2.0x10$^6$ | 2.0x10$^{-8}$ |
| B | 14 | 0.18 | 1.1x10$^{62}$ | 1.3x10$^6$ | 8.1x10$^{-9}$ |
| C+G | 16 | 0.13 | 2.8x10$^{62}$ | 3.8x10$^6$ | 6.7x10$^{-9}$ |

Table5. The calculated flux of non-thermal bremsstrahlung and gamma ray flux from pion decay at 1 GeV.

| Name | $N+n_i$ (cm$^{-3}$) | $K'$ (m$^{-3}$GeV$^{p-1}$) | $p$ | Non thremal bremsstrahlung flux at 1 GeV (ph/s/cm$^2$/GeV) | pion decay flux at 1 GeV (ph/s/cm$^2$/GeV) |
|---|---|---|---|---|---|
| N | 2.1 | 7.1x10$^4$ | 3.5 | 8.4x10$^{-14}$ | 1.1x10$^{-19}$ |
| HST-1 | 0.6 | 3.0x10$^{11}$ | 3.5 | 1.0x10$^{-7}$ | 1.4x10$^{-13}$ |
| D | 203 | 1.5x10$^{11}$ | 4 | 4.2x10$^{-2}$ | 2.4x10$^{-11}$ |
| A | 42 | 4.4x10$^{11}$ | 3.5 | 1.0x10$^{-5}$ | 1.4x10$^{-11}$ |
| B | 17 | 9.0x10$^{10}$ | 3.5 | 8.6x10$^{-7}$ | 1.2x10$^{-12}$ |
| C+G | 20 | 5.6x10$^{10}$ | 3.5 | 6.3x10$^{-7}$ | 8.7x10$^{-13}$ |

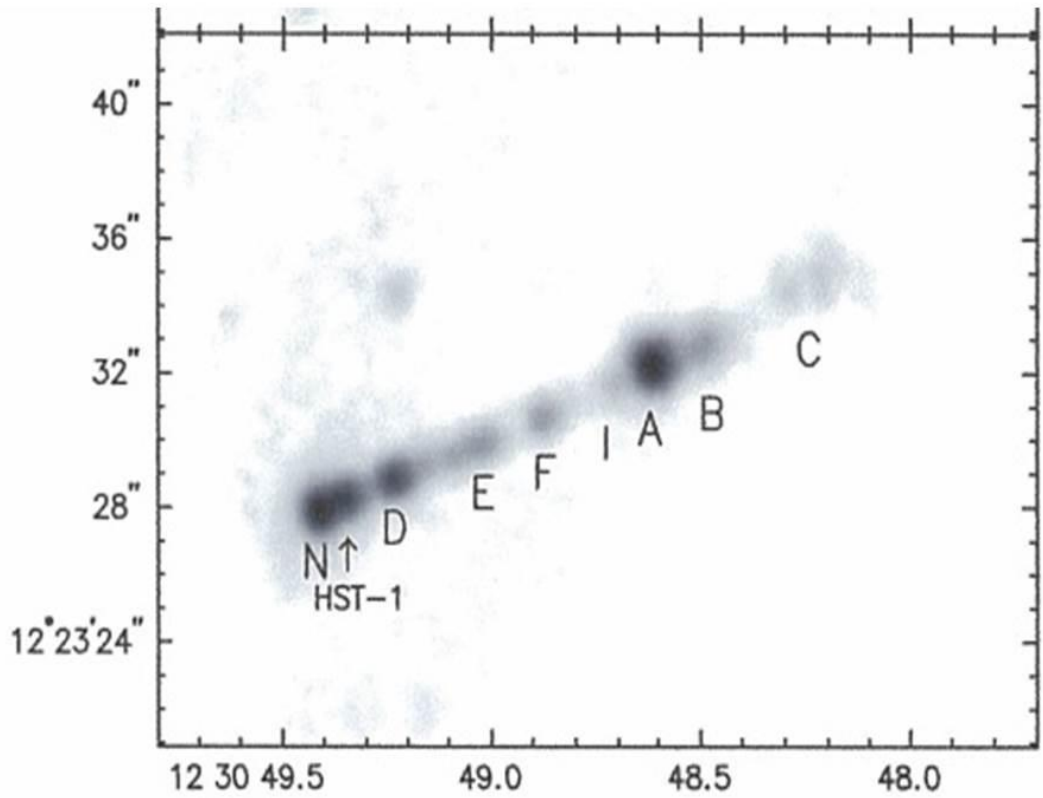

Figure1. X-ray image of the M87 jet with *Chandra* extracted from Wilson & Yang (2002). X-axis is RA and Y-axis is DEC.

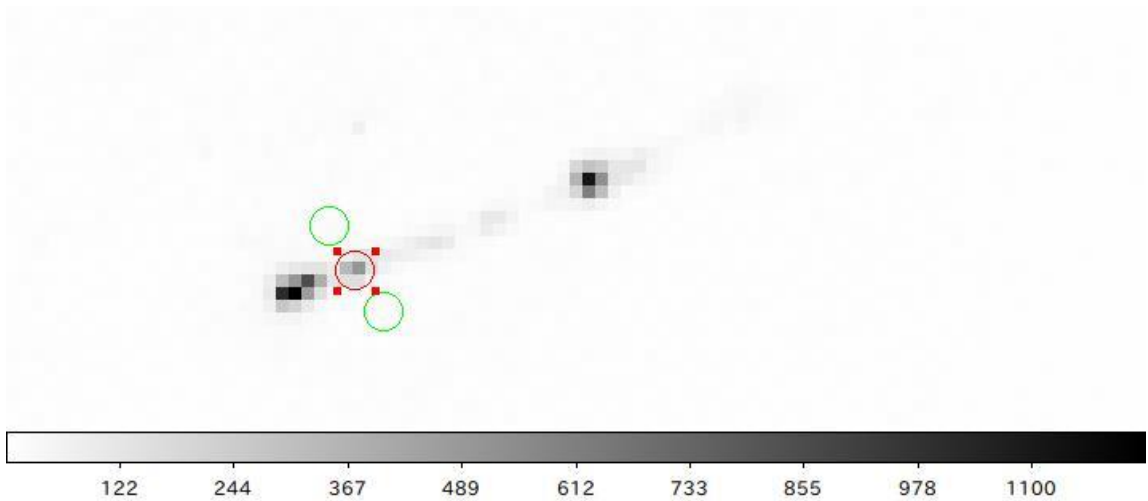

Figure2. The source region (red circle) and two background regions (green circle) for field D. Grey scale is intensity.

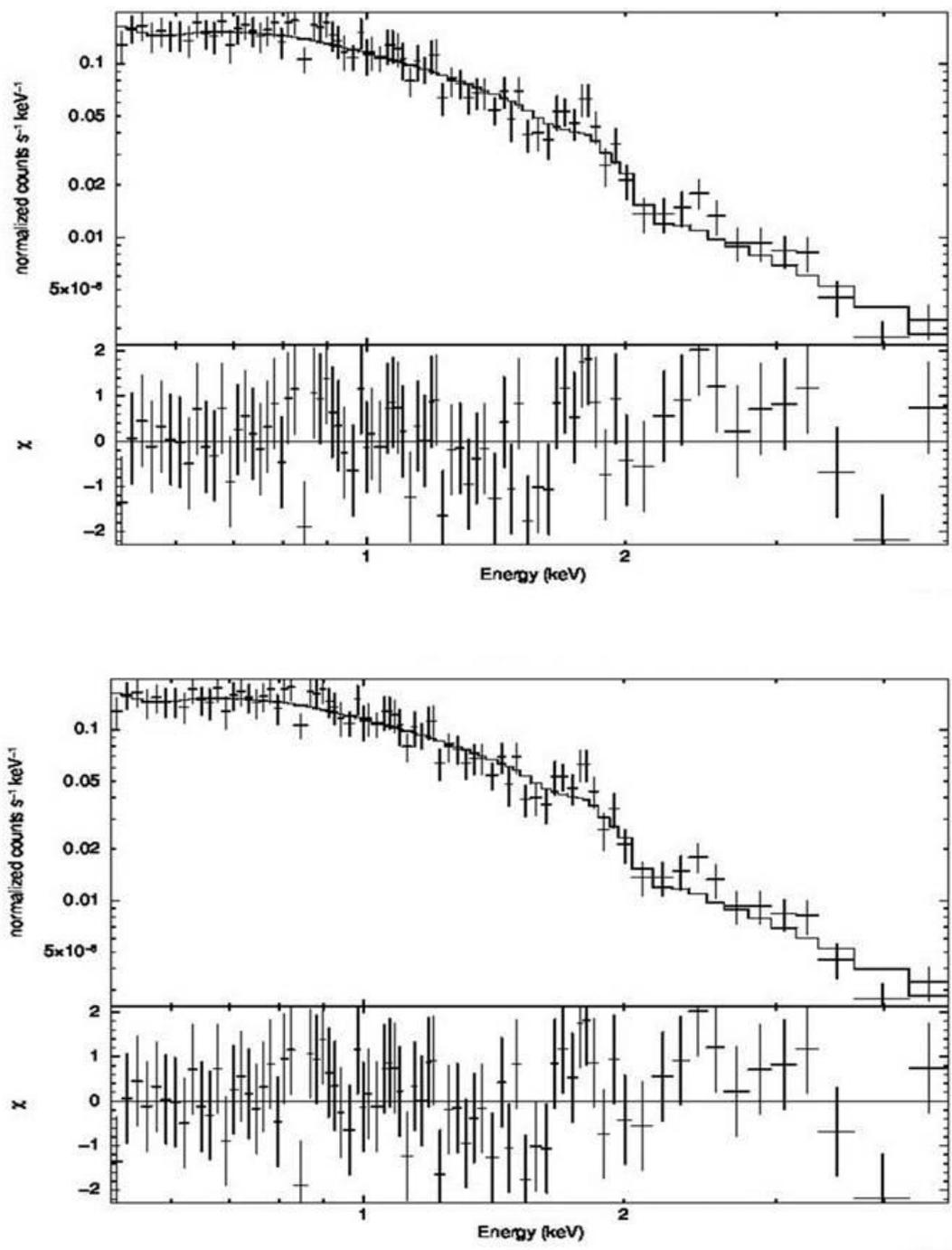

Figure3. The fitted X-ray energy spectra for the field N. The models used are power law(top) and a combination of a power law and thermal bremsstrahlung (bottom). Photo absorption is included.

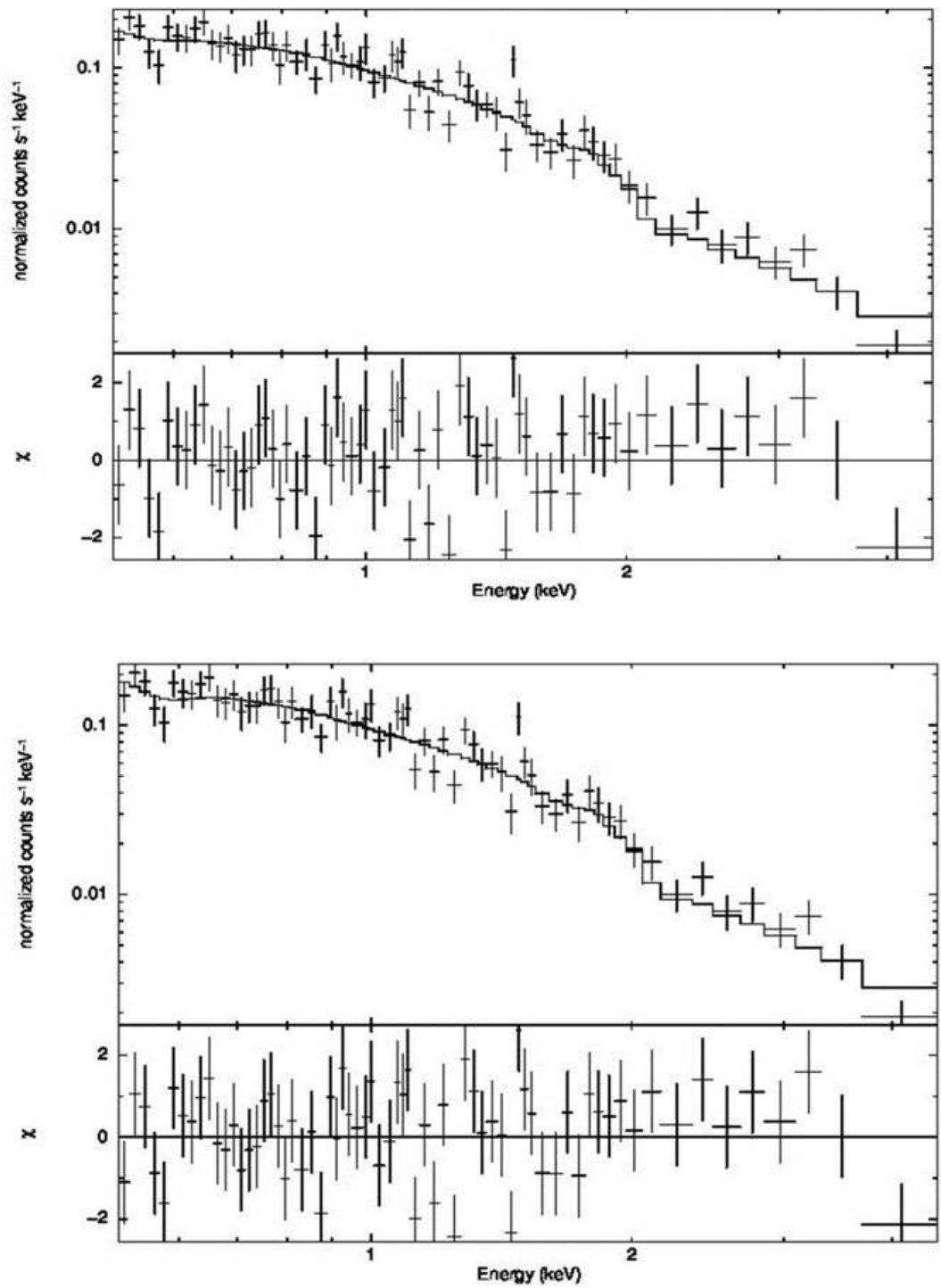

Figure4. The fitted X-ray energy spectra for the field HST-1. The explanation is same with Figure3.

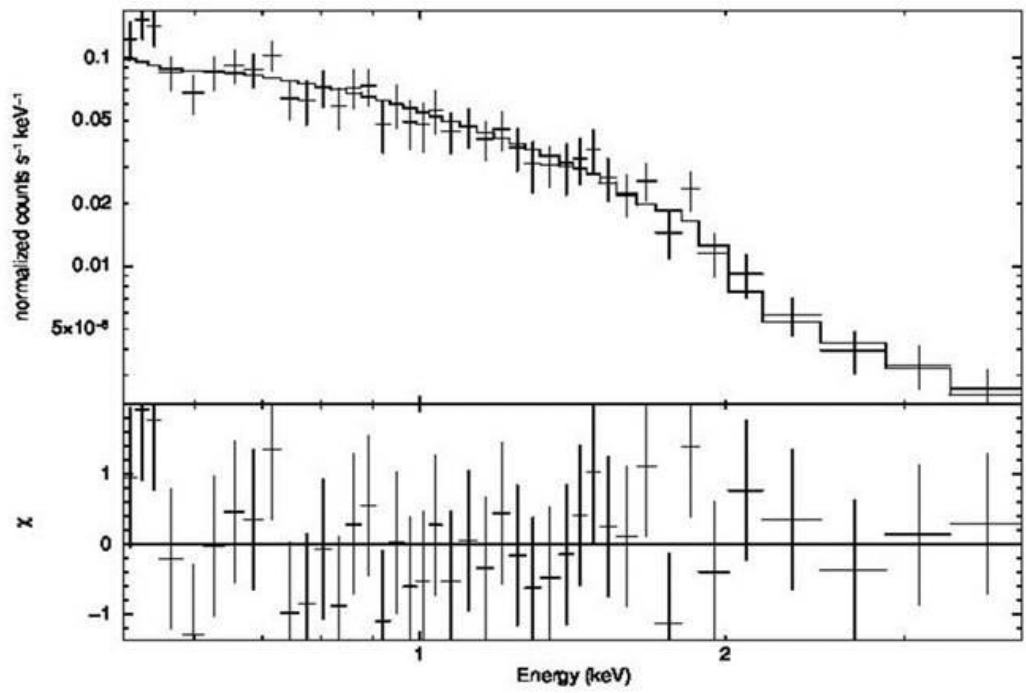

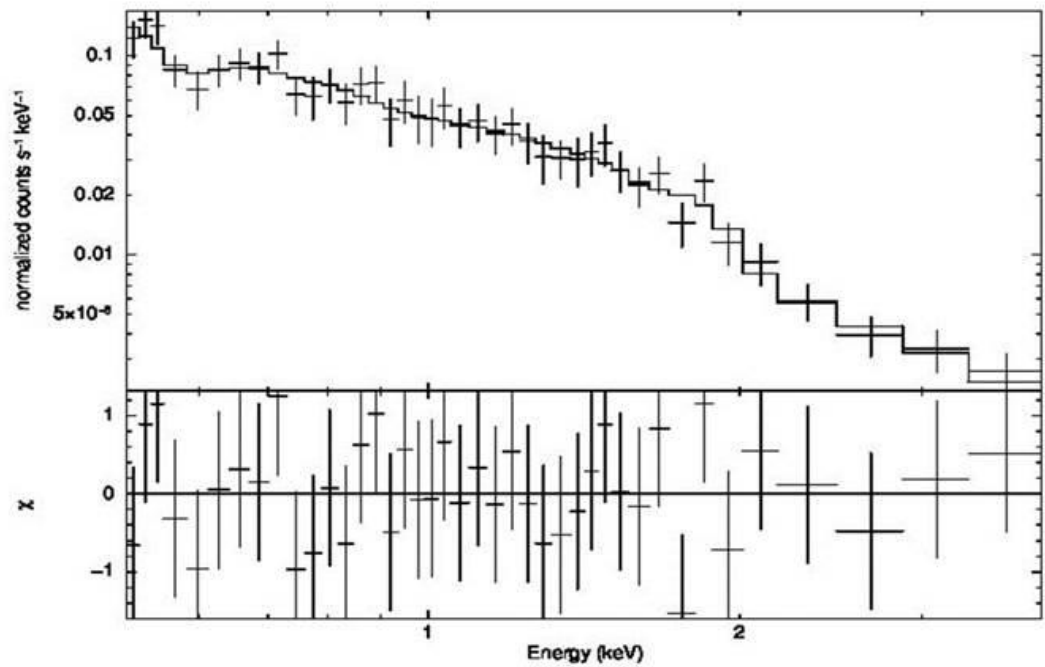

Figure5. The fitted X-ray energy spectra for the field D. same description with Figure4.

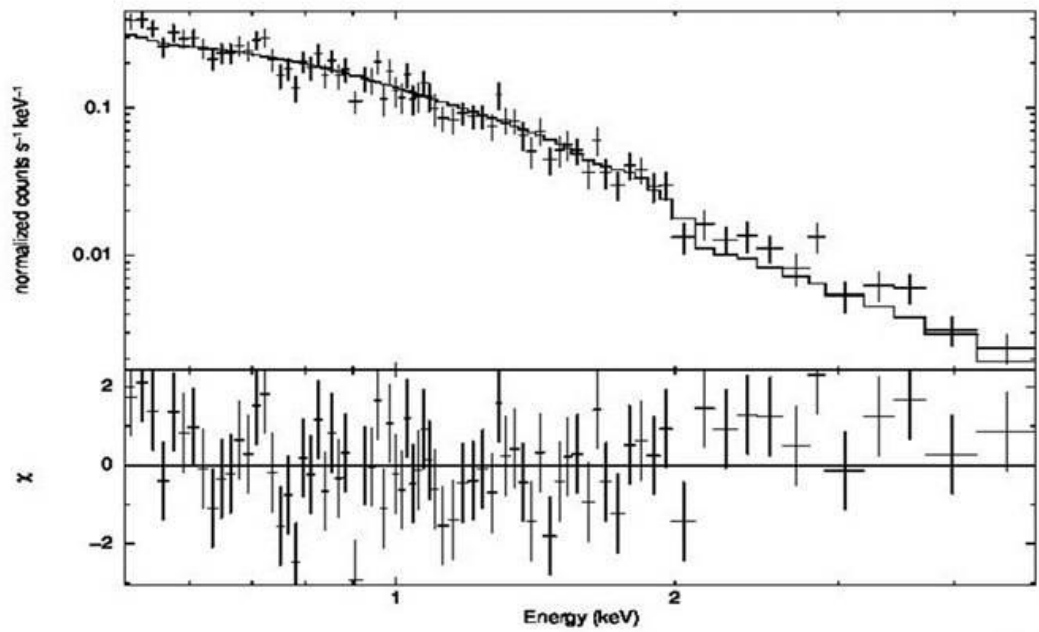

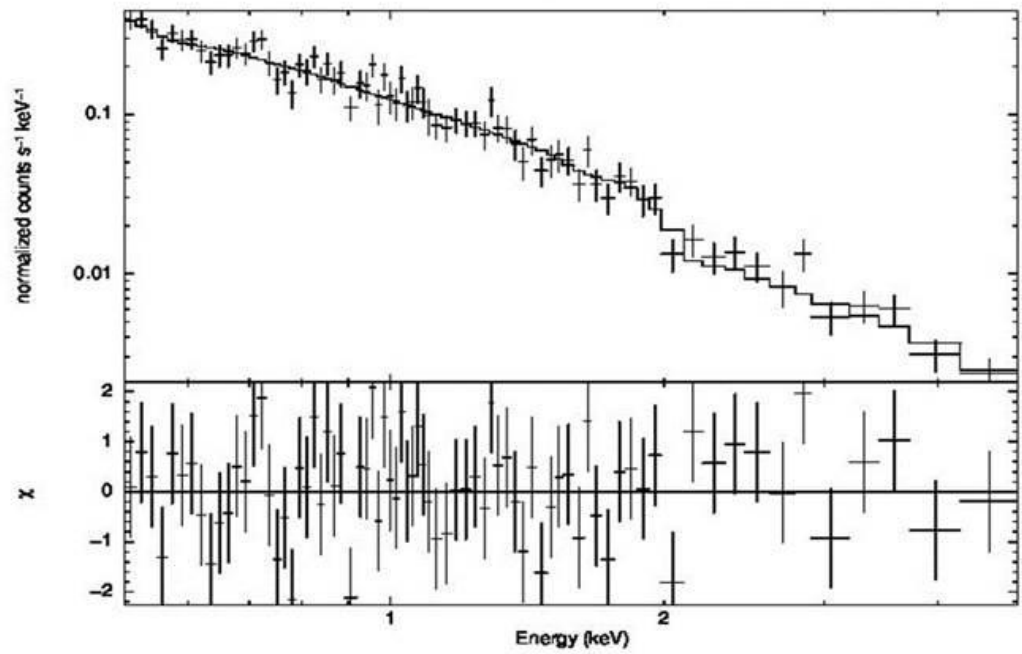

Figure6 The fitted X-ray energy spectra for the field A. The explanation is same with Figure3.

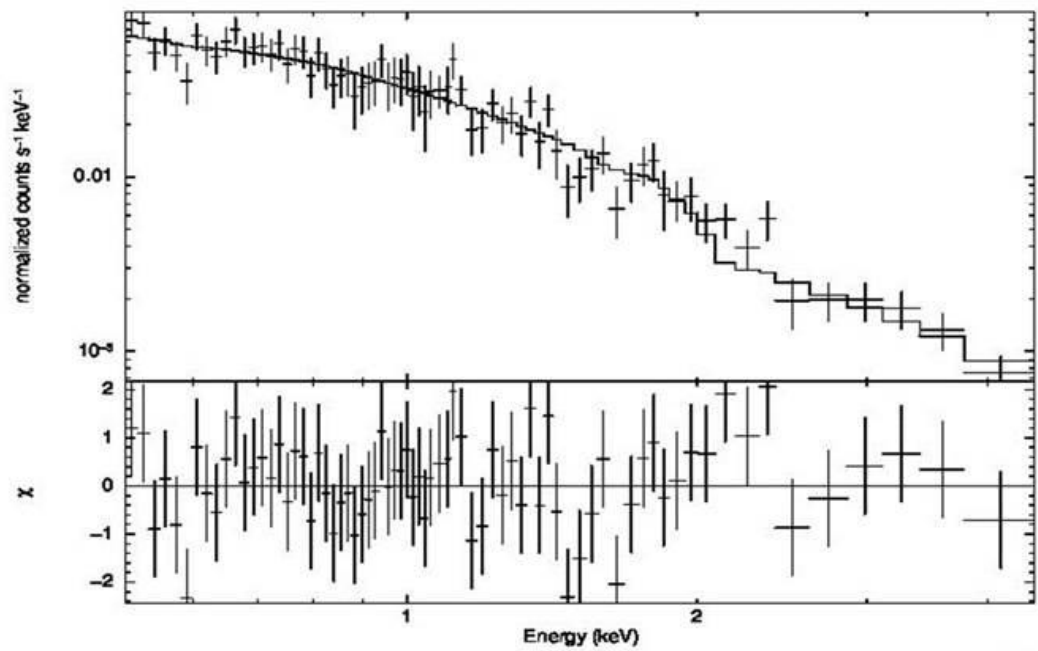

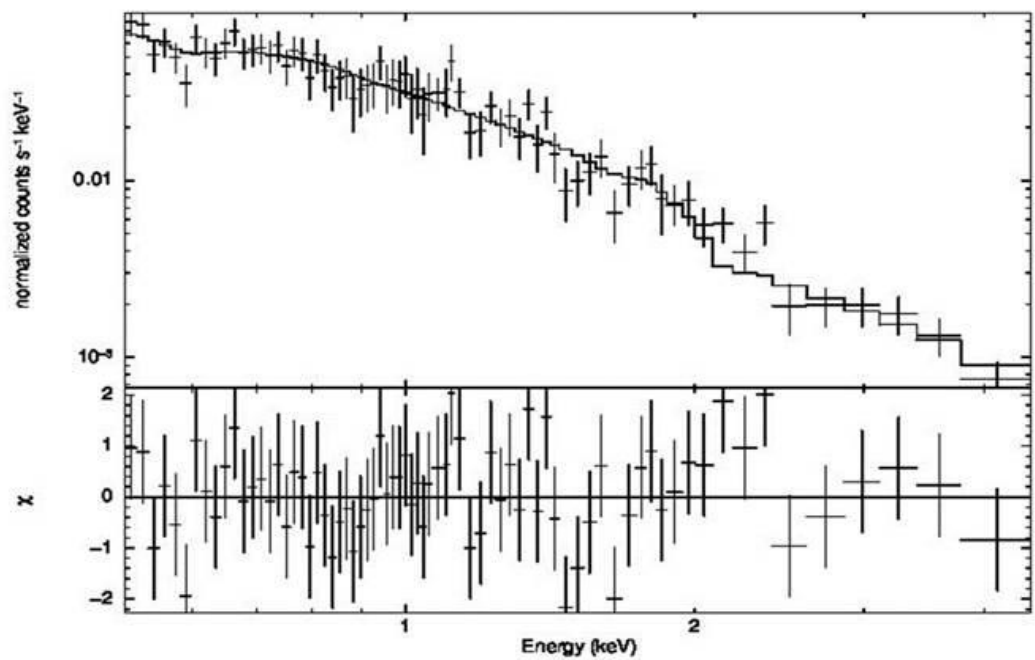

Figure7 F The fitted X-ray energy spectra for the field B. The explanation is same with Figure3.

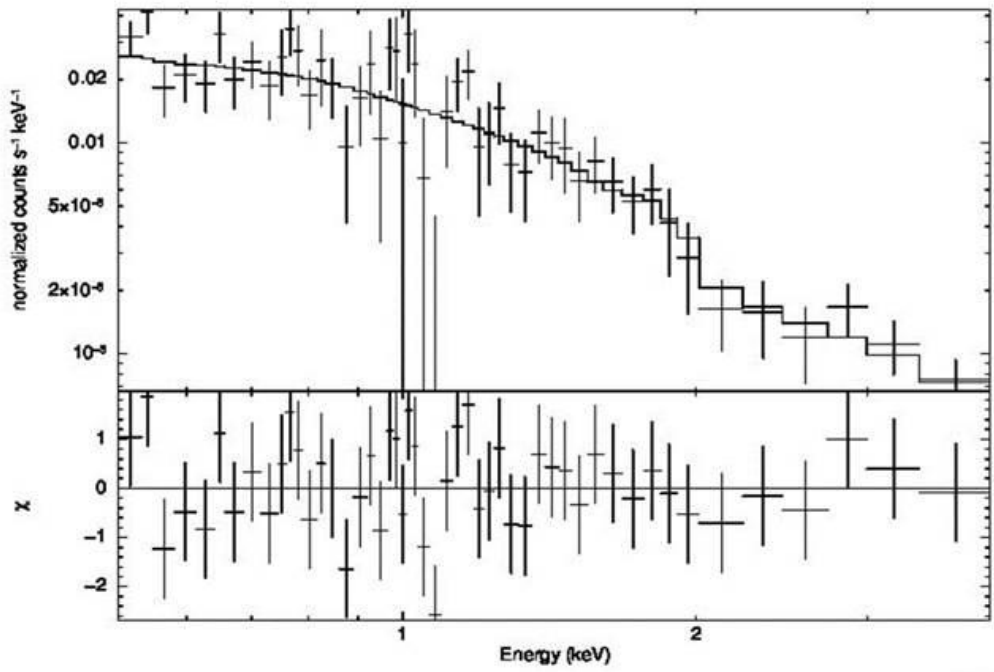

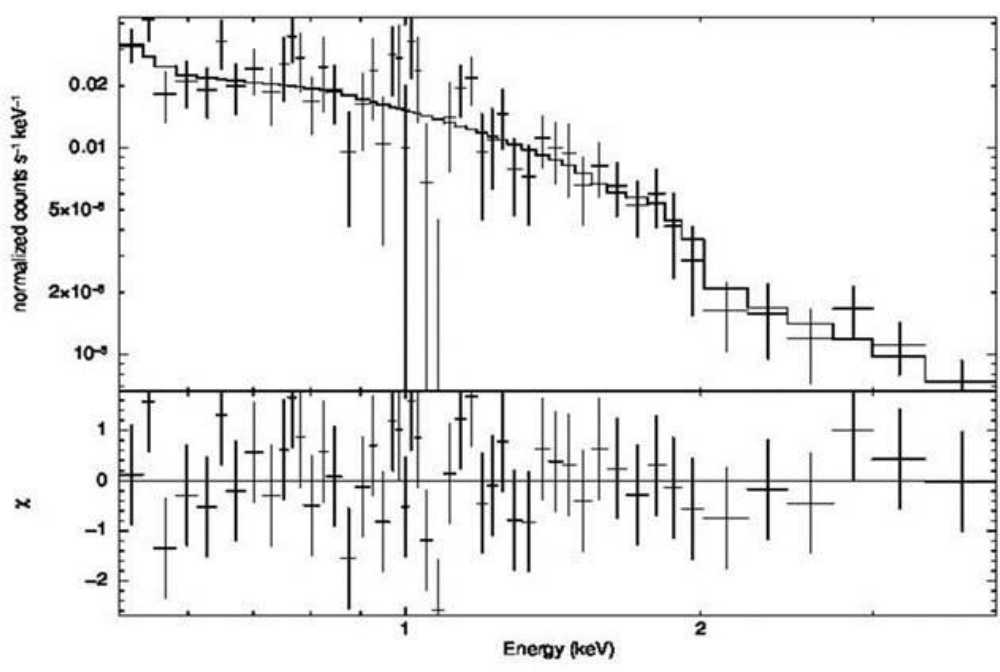

Figure8 The fitted X-ray energy spectra for the field C+G. The explanation is same with Figure3.